# Opto – acoustical gravitational bar detector with cryogenic mirrors

V.V.Kulagin[1], S.I.Oreshkin[1], S.M.Popov[1], V.N.Rudenko[1], M.N.Skvortsov[2], I.S.Yudin[1]

[1] Sternberg Astronomical Institute of Moscow State University
[2] Institute of Laser Physics, Siberian Branch of RAS

**Abstract**. Enhancing of sensitivity of the opto-acoustical gravitational wave (GW) antenna OGRAN installed in the underground facilities of Baksan Neutrino Observatory is analyzed. Calculations are presented showing a sensitivity improving on two orders of value after a cooling the solid body acoustical part of the antenna to the nitrogen temperature. A possibility of keeping of the same optical scheme of the antenna at low temperature is discussed. Design of modernized construction for cryogenic version of the antenna OGRAN is described. Test experiments with cooled pilot model carrying cryogenic mirrors illuminated by the optical pump up to 0.5 W are presented.

## 1. Introduction.

Gravitational antenna OGRAN is the setup having a combination of acoustical and optical principles of gravitational wave detection. OGRAN was developed and constructed by collaboration of Moscow State University (Sternberg Astronomical Institute, SAI MSU) and Russian Academy of Sciences (Institute of Nuclear Research, INR RAS, and Institute of Laser Physics, ILP SB RAS). Detailed description of this antenna was given in [1,2]. At present the antenna OGRAN is installed in underground facilities of the Baksan Neutrino Observatory of INR RAS and is going on through a commission stage. Using this instrument, a long time observation of the gravity gradient background is planned in parallel with neutrino events monitoring at the neutrino telescope setup (BUST) [3] having the goal of a joint search for collapsing stars – relativistic transient events in our Galaxy and close halo region with radius ~100 Kpc.

Extension of the zone of detectable sources requires an instrument with enhanced sensitivity. For the OGRAN antenna construction, it could be achieved in particular using its cryogenic version with cooled solid body GW detector. Although a technique for resonance bar cryogenic gravitational wave detectors has been well developed and tested in long observational runs [4,5], it can't be applied directly to the OGRAN antenna because it meets the problem of realization of optics with cooled mirrors functioning under enough power optical pump. It is worth to remark here that the same problem arises in a development of laser interferometer GW detectors of the third generation, so called "Einstein Telescope", in which mirrors being under the helium temperature have to operate with the optical power on the order of hundred watts [6].

The objective of this paper consists in the analysis of achievable sensitivity of the combined opto-acoustical antenna OGRAN under a moderate but reasonable cooling of its acoustical part keeping the same design for optical parts. It is clear that the current technological state of art will limit the optimal optical pump power under which such antenna could operate in a duty cycle of a long time continuous observation.

## 2. OGRAN construction and operation principle.

OGRAN gravitational antenna construction is based on a classical scheme known as the "comparator of EM-frequency etalons" in which the detection signal arises as a result of comparison of a couple of optical high quality resonators (to clarify see Fig 3 in section 5 below). An external stable laser is used for the common optical pump but its frequency is referred to the one, large, resonator which is considered as the gravitational detector (or GW receiver). Mirrors of this resonator attached to the ends of a large solid cylindrical acoustical bar (length of $L=2$ m and mass of $M \sim 2$ t) with longitudinal axial tunnel compose a high finesse Fabry-Perot (FP) cavity. Gravitational wave produces a perturbation of both acoustical and optical degrees of freedom shifting the optical resonance. A detuning of the GW detector cavity is registered by differential comparator circuit including a second short stable FP cavity ($l=15$ cm), operating as the optical discriminator. For the thermal stability, the body of discriminator is composed from Sitall CO-115M – material with extremely small thermal expansion coefficient [7]. Output signal of this scheme is proportional to the frequency difference of laser and discriminator. Using the opto-electronic feed back circuits this difference is reduced to zero. As a result, the laser beam frequency carries information about variations of the GW detector optical length (bar end displacements or refractive index changes).

Initial working regime of the setup or the "antenna's operating point" corresponds to precise resonance tuning of both cavities. For the detector cavity, it is attained by variation of the pump laser frequency. The discriminator cavity has one of its mirrors attached at a piezoelectric tablet driven by a voltage of feedback circuit. By this way the cavity resonance position at low frequencies (less the 100 Hz) is fixed. At the signal frequencies close to 1 KHz, the discriminator cavity is free. Error signals in both cavities are produced by the well known Pound-Drever-Hall method which applies the radio frequency (on the order of 10 MHz) phase modulation on the input optical beam [8]. The beam reflected from the discriminator cavity is synchronously demodulated at the photo detector. The photo current is proportional to the difference between the laser frequency and the Eigen discriminator frequency, which is supposed to be stable enough. So the output signal reflects only displacements of the GW detector cavity. For the sustainable operation of the whole antenna it is necessary to stabilize also the amplitude of the laser pump.

The optical scheme of OGRAN has to operate in the regime of "zero output signal" when external gravitational perturbations are absent, so called "the work in the dark spot (fringe)", as it is accepted in all long based gravitational interferometers. At the same time it does not require a direct interference between two optical beams that essentially facilitate the technical realization of the setup.

### 3. Sensitivity, noise factor and reception frequency bandwidth.

The unavoidable noise background for the OGRAN setup is composed by the two natural sources of fluctuations: thermal noise of the acoustical bar and photon noise of the optical read out. Effectiveness of these sources depends on the key parameters of the setup such as mechanical losses, temperature, optical power, cavity finesse. Below in this section, we present a simple analysis with the goal to define optimal combination of different antenna parameters which allows increasing the OGRAN sensitivity without unrealistic complication of the setup and high cost expenses.

The potential sensitivity of a resonance bar gravitational detector defined by its Brownian fluctuations is described by the following formula [9, 10]

$$h_{\min} = \frac{4}{L}\left(\frac{kT}{M\omega_\mu^2}\right)^{1/2} \frac{1}{[\omega_\mu Q_\mu \tau]^{1/2}} F^{1/2}, \qquad (1)$$

where $L$, $M$, $\omega_\mu$, $Q_\mu$ are length, mass, resonance frequency and acoustic quality factor of bar detector, $k$ is the Boltzmann constant, $F$ - is a noise factor for setup, $T$ - is bar temperature, $\tau$ - is GW burst duration. For $M = 10^3$ kg, $Q_\mu = 3 \cdot 10^5$, $\omega_\mu = 10^4$ s$^{-1}$, $\tau = (1/\Delta f) = 10^{-3}$ s, $T = 300$ K, (as the mass we used the reduced equivalent – half of the bar mass) the sensitivity estimation according to (1) results in

$$h_{\min} \approx 10^{-20}(F\Delta f)^{1/2} \qquad (2)$$

One can remark at once that cooling to the liquid nitrogen temperature $T=77$ K decreases the thermal noise spectral density down to $2.5 \cdot 10^{-21}$ Hz$^{-1/2}$ (for $F\Delta f \sim 1$). Meanwhile under cooling the quality factor can to be increased at least on one order of value, i.e. $Q_\mu = 3 \cdot 10^6$. It allows expecting the sensitivity limited by the bar thermal noise at the level $\sim 10^{-21}$ Hz$^{-1/2}$. The noise factor $F$ defines the excess of the setup real noise level above the variance of the thermal noise. The formula for the noise factor of optical displacement sensor read as [9,10]

$$F = [\delta\tau \cdot \arctan(1/\delta\tau)]^{-1}, \qquad (3)$$

Where $\delta^2 = \delta_\mu^2 + (2M)^{-2} \cdot G_B/G_\nu \gg \delta_\mu^2$, $\delta_\mu = \omega_\mu/(2Q_\mu)$,

$G_B$ is a spectral density of the thermal noise of the bar detector and $G_\nu$ is a spectral density of an additive optical noise of the registration system:

$$G_B = 2kTM\omega_\mu/Q_\mu, \qquad G_\nu = B\omega_\mu^2 (2h\nu/\eta P)(\lambda_e/2\pi N)^2 \qquad (4)$$

$\omega_e = 2\pi\nu$ and $\lambda_e$ are the frequency and the wavelength of the optical pump, $P$ is a pump laser power, $\eta$ is a quantum efficiency of the photo detector, $N$ is a number of reflections in the FP cavities, and factor $B$ characterizes an excess of the real laser noises above the short noise level of ideal optical pump (i.e. part of technical noise above the Poisson quantum noise). We do not take into account the "back-action noise" of the optical read out because it is much less than the Brownian nose for the moderate pump power used in setup.

For numerical calculation of optimal combination of variable parameters of the cooled antenna, we will use below the following (typical for OGRAN) constant values: $L=2$ m, $M \cong 2000$ kg, $Q_\mu = 2 \cdot 10^6$ (considering that acoustical losses decreased under cooling), $\omega_\mu = 10^4$ s, $\tau = 10^{-3}$ s. The temperature $T$ in our calculation will be considered as the variable argument inside the interval $(3 \div 300)$ K. For the optical readout system, we will suppose the following data: $P=100$ mW $\div$ 1 W, $N = 10^4 \div 10^5$, $\eta=0.7$, $B=1 \div 10$, $\lambda_e =1$ μm. It is worth to remark that a high stable laser with the power of 100 mW ÷ 1 W and FP cavity with the number of reflections (finesse) equal to $10^5$ are available in modern optical labs and present enough routine experimental technique.

Numerical calculation with formulae (1) – (4) results in characteristics of the cryo-OGRAN antenna presented in Fig. 1, which depend on the level of bar detector cooling and excess noise of the optical readout.

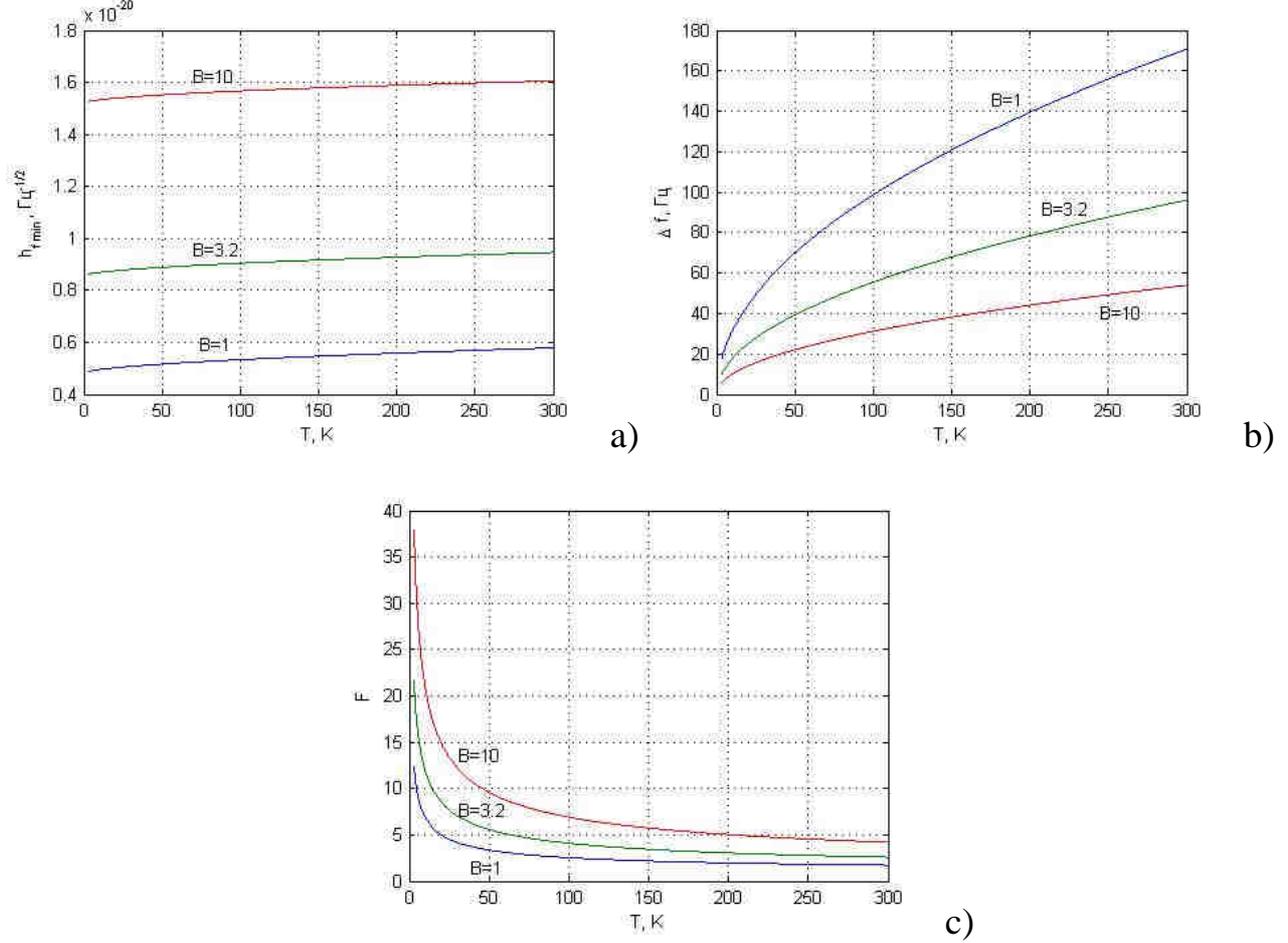

Fig. 1. Sensitivity characteristics of the cryo-OGRAN antenna: a) spectral sensitivity; b) bandwidth of the optimal filtration; c) noise factor F (P=1 W, N=$10^4$), parameter of the laser excess noise is: B=1 (blue curve), B=3.2 (green curve), B=10 (red curve).

One can see from Fig. 1 that at temperatures 50 – 100 K, the optical read out has the low noise factor, i.e., it is practically an "ideal registration system" without an additive noise. At the same time, the receiver bandwidth is several tens of Hz, and the amplitude spectral noise density $h_{min}$ is only few times larger than the level of potential sensitivity with ideal registration system $h_{min\,pot} \sim 10^{-21}$ Hz$^{-1/2}$. It is clear also that the sensitivity slightly changes with variations of the main parameters of the optical system that provides a possibility for a stable operation during the long time observations.

### 4. OGRAN read out arm and influence of discriminator noises.

The "read out" arm of the OGRAN setup containing the discriminator cavity remains under room temperature. In general, a variation of the optical resonator length $l$ leads to the change of the resonance frequency ν according to the simple relation $\Delta\nu = (\Delta l/l)\cdot\nu$, i.e., the effect is stronger for the shorter resonator (in the OGRAN setup, $L$=2 m and $l$=50 cm [1,2]). On the other hand, a discrimination of the frequency shifts will be better with a narrow resonator bandwidth. For this purpose it is necessary to use a long cavity. The contradiction can be resolved for a short resonator - discriminator with enough high finesse.

It was mentioned that the pump laser frequency is coupled with the resonance frequency of the large FP cavity fixed at the GW acoustical bar detector. Accuracy of such coupling in the ideal case is limited by the quantum photon noise and equals to the standard $\Delta\nu = (c/4LF)(h\nu/\eta P)^{1/2}$ with dimension [$Hz/Hz^{1/2}$]. The photon shot noise defines also the minimal registered spectral amplitude of GW bar detector length variation $\Delta L=(c/4\nu F)(h\nu/\eta P)^{1/2}$ with dimension [$m/Hz^{1/2}$]. It has to be measured at the discriminator output. For precise measurement, the discriminator finesse has to be larger than the finesse of GW bar detector FP cavity. However, the thermal noise of the discriminator can limit the registered amplitude of the GW detection in accordance with the formulas written above (see section 3).

To decrease contribution of the discriminator thermal fluctuations, the mirrors attached to it's ends should have acoustical resonance frequency much higher than the fundamental acoustical frequency of GW bar detector. The geometrical length of discriminator in the OGRAN setup was taken as 50 cm (four times less the bar scale). So the operational detection frequency 1.3 KHz was occurred to be lower enough than the discriminator longitudinal mode frequencies ($\geq$ 5 KHz). In this case, the spectral density of thermal oscillations of the discriminator end mirrors $x_m$ is read as

$$x_m^2(\omega) = \frac{1}{(\omega_d^2 - \omega^2)^2 + 4\delta_d^2\omega^2} \cdot \frac{2kT\omega_d}{\pi m_d Q_d} \approx \frac{2kT}{\pi m_d \omega_d^3 Q_d}, \quad (5)$$

where the down index $d$ is referred to discriminator parameters.

Thus, the thermal noise of the discriminator can be considered as a white noise in the area of operation frequency as well as optical photon noise. It allows describing these noises by introducing the effective extra noise coefficient $B_{eff}$ with respect to the noise background of the large GW detector. Reducing these extra noises to the input of the OGRAN setup, one comes to the following formula for $B_{eff}$

$$B_{eff} = 1 + \left(\frac{LN}{lN_d}\right)^2 \frac{P}{P_d} + \frac{\left(\frac{L}{l}\right)^2 x_m^2}{\left(\frac{\lambda}{4\pi N}\right)^2 \frac{\hbar\omega_e}{\eta P}}, \quad (6)$$

here $l$, $N_d$, and $P_d$ are for the length, the number of reflection and the optical power of the discriminator. To keep the setup sensitivity close to the potential one (Eq. (1)), it is necessary to minimize $B_{eff}$. One approach for this is the selection of the discriminator mirrors according to the relations $LN \cong lN_d$ and $P \cong P_d$. According this equation all parameters of the discriminator are defined and it is possible to calculate the value of $B_{eff}$. The corresponding graphs versus of temperature and different parameters of the setup are given in Fig 2.

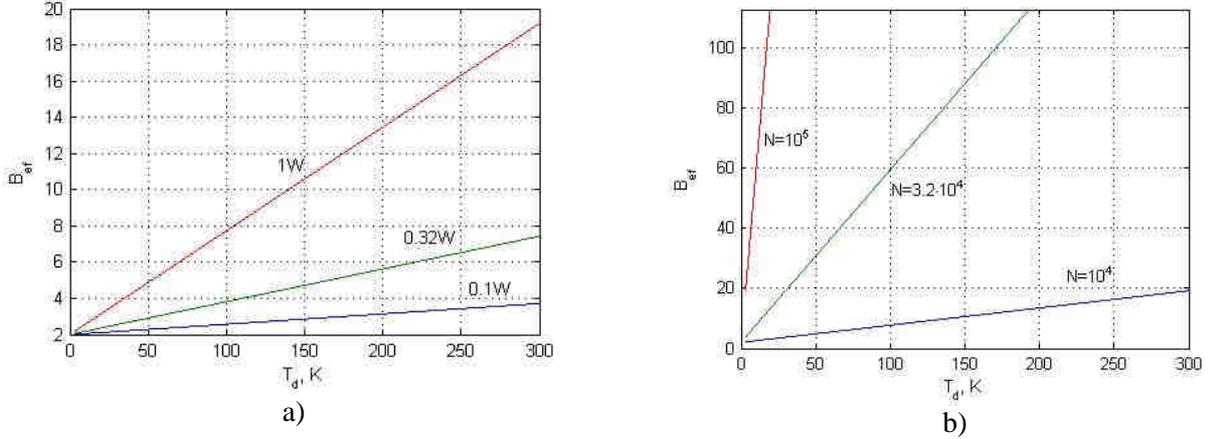

a)            b)

Fig. 2. Excess noise coefficient $B_{eff}$ versus cooling temperature for different parameters: a) laser powers and b) numbers of reflections in the FP cavities.

It is possible to conclude that the discriminator noises essentially decrease the OGRAN sensitivity only for extreme parameters of the optical readout when the optical sensor is close to the ideal one. A decreasing is remarkable for the high temperature with the maximal filtration bandwidth. At the moderate (realistic) parameters of the registration arm, the value $B_{eff}$ is close to 10, that decreases the sensitivity by 2-3 times only (Fig. 1a).

**5. OGRAN test setup with cryogenic mirrors.**

In the process of development of the OGRAN antenna, many technical and technological elements were invented and tested with its test setup [11]. Modernized version of this setup was implemented for the development of the OGRAN cryogenic setup. Also, a special helium cryostat was used for testing mirror's parameters evolution during the cooling process [12]. Optical scheme of the modernized test model is presented in Fig. 3.

Reception characteristics of the antenna in test setup will be defined by its parameters: mass $m=50$ kg, length $l=50$ cm, acoustical quality factor $Q_\mu = 10^4$, and resonance frequency $\omega_\mu = 4 \cdot 10^4$ s$^{-1}$. The sensitivity estimation can be found through the same formulas as for the large antenna (1) - (4). Analysis shows that for the test setup, parameters of the optical sensor can be taken not so extreme as for the large antenna, namely: the pump power can be 0.1 W and the number of reflection can be $N=10^3$. The antenna in test setup cooled down to the liquid nitrogen temperature will have the sensitivity $\sim 10^{-18}$ (Hz)$^{-1/2}$ in the "strain terminology", i.e. by two orders of value better than the room temperature version [11]. Under this conditions, the optical registration system occurs to be very close to the ideal sensor (the sensitivity is worse than the potential one only by 2-3 times), and the thermal noise of the discriminator practically does not produce any influence on the total setup sensitivity.

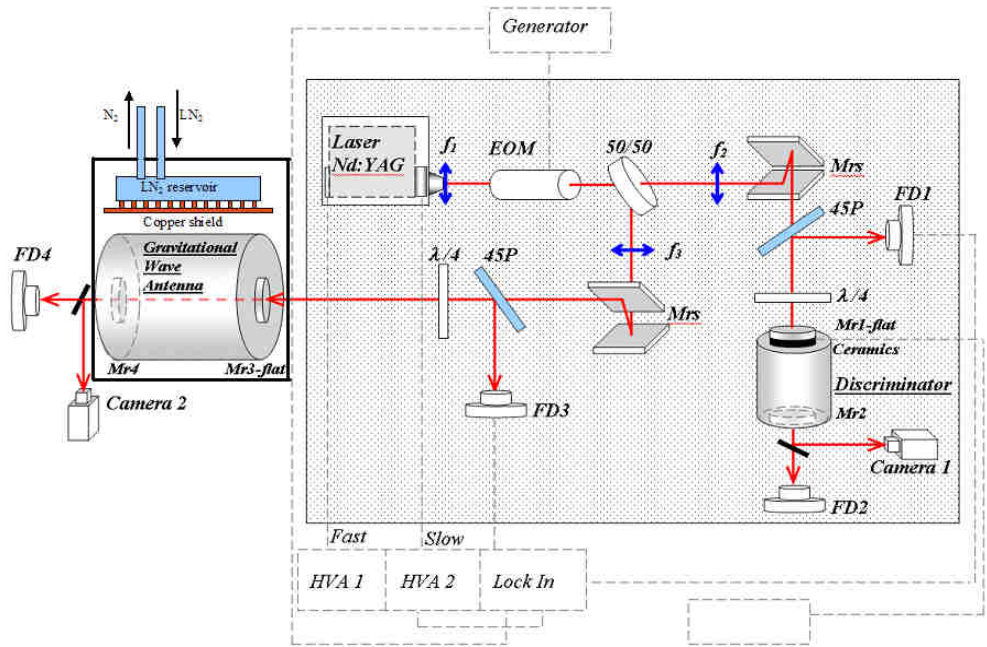

Fig.3 Opto-mechanical scheme of the cryogenic OGRAN test setup

Schematic picture of the vacuum cryogenic chamber (1) with a simple wire anti-seismic suspension of the test setup of the acoustical resonance bar (2) is presented in Fig. 4.

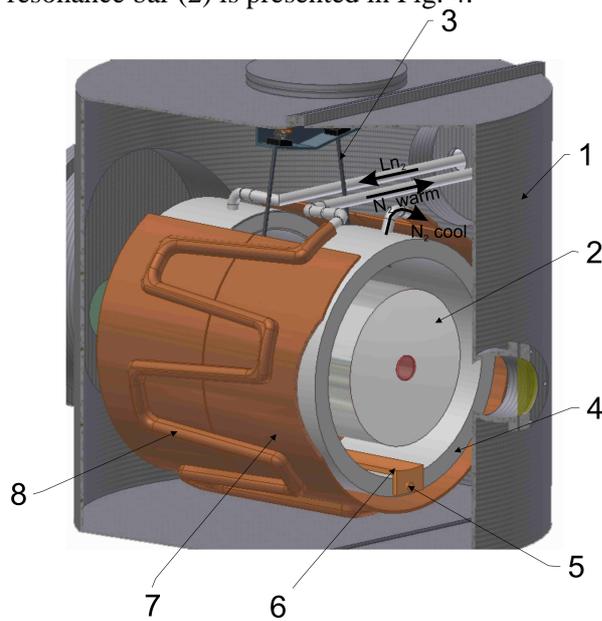

Fig.4. Schematic picture of the vacuum cryogenic chamber: 1 - main vacuum chamber, 2 – detector with mirrors, 3 - stainless wire suspension, 4 - *LN2* reservoir, 5 - copper rod, 6 - copper plate, 7 - copper radiation shield, 8 - copper recuperation tube.

The test setup is suspended through a stainless wire loop (3) in the vacuum chamber inside the concentric liquid nitrogen ($LN_2$) reservoir (4). Two thick enough (12 mm in diameter) copper rods (5) are soldered with high-silver content brazing alloy into the $LN_2$ reservoir in its lower part. In its middle part, the pilot model is surrounded by the thick enough (8 mm) copper plate (6) attached to these rods with a good thermal contact. The shape of the copper plate allows one to connect it with the pilot model by several soft copper-wire thermal links to achieve a good thermal contact between the $LN_2$ reservoir and the pilot model. The $LN_2$ reservoir and the pilot model are completely surrounded with the copper radiation shield (7) in a vacuum environment. A copper tube (8) with a loop form is soldered along the total length of the radiation shield. The temperature of the radiation shield decreasing sufficiently by driving the evaporated gas from the $LN_2$ reservoir through that copper tube. The $LN_2$ coolant and escaping gas flow are shown in Fig. 4 by arrows.

There are very strict requirements to the level of vacuum inside the chamber in such configuration. Calculation shows that for minimization of the heat transfer from the chamber walls ($T_2$=300 K) to the pilot model body ($T_1$= 77 K) by residual gas, the vacuum must be no worse than $5 \cdot 10^{-4}$ Torr. Then, the heat transfer will be conducted mostly by electromagnetic radiation and can be estimated as

$$W = (\Delta Q / \Delta t) = \varepsilon_n \sigma (T_2^4 - T_1^4) A_{in} , \qquad (7)$$

where $\sigma = 5.67 \cdot 10^{-12} \ W/cm^2 \ K^4$ is the black body radiation constant, $\varepsilon_n \sim 0.01$ is the emissivity and $A_{in}$ is detector's surface area of the test setup. In our case, the radiation flux will be locked at the radiation shield and affects only the value of the nitrogen expenses in the cooling process.

Calculation shows that approximately 200 hours are required for cooling down the model bar from 300 K to 78 K only through the radiation flow. For to reduced this time four soft copper wire conductors attached by screws to the bar surface around its central circle and locked to the cryo-panel were foreseen. In this configuration, the estimated cooling time is 1.5 hours only for one cooling conductor with the length of 10 cm, diameter of 0.5 cm and thermal conductivity mean value coefficient in the copper – aluminum contact of $\lambda_{contact}$= 1 W/cm K.

### 6. Technology of cryogenic optics.

One of the large technological problems for the cryogenic version of the OGRAN setup is to keep mirror's optical characteristic invariable at low temperature under illumination of the 0.1-1 W laser power. At such power level, effect of the thermal induced lens of the mirror can have significant influence on the optical FP mode structure. The mirror's substrates also should have low thermal noises not to depress the high strain sensitivity and good enough thermal conductivity at low temperatures to dissipate absorbed power. At the present, $CaF_2$, sapphire and $Si$ mono crystal are considered as promising materials for cryogenic mirror's substrates [13]. The *Sapphire* and $CaF_2$ have a broad transmission spectral range but the first is relatively expensive while the second has some problem with quality polishing. $Si$ single crystals can be obtained in large volumes without structural defects having a very low impurity level. However the OGRAN setup uses Nd:YAG single frequency laser operating at 1064 nm. For this wavelength $Si$ is not transparent and so it can't be a proper choice for OGRAN cryo-version. With this argumentation the $CaF_2$ substrates were selected in our test experiments to study of optical characteristics of cryogenic mirrors during the cooling process with the presence of optical pump.

Two mirrors on $CaF_2$ substrate were attached with a good thermal contact to the ends of small model of the OGRAN detector of cylindrical shape with 14 cm diameter and 20 cm length made from aluminum alloy (total weight 8 kg). The cylindrical body had a 2 cm hole along main axis to transmit laser light in FP cavity. The aluminum body with attached mirrors was placed in the special cryostat [12] also in the good thermal contact with walls of cryostat inner chamber. To minimize losses of thermal irradiation, the inner chamber was covered by several screens with liquid helium and nitrogen temperatures. The only one optical window with 2 cm in diameter was foreseen at the end cover of the cryostat together with co-centric holes in the inner envelopes. Through this window the light of pump laser can illuminate the FP resonator associated with the model detector (details of the cryostat and the model one can take in [12]). Reflected from FP resonator light might be picked up with outside photo detector.

The goal of our experiments was consisted to measure FP cavity integral optical characteristics during the cooling process from room temperature up to liquid helium temperature and then during the back process of warming to room temperature. It took more than a week for the whole thermal cycle. For the actual mirror temperature control we used two thermo sensors: one on the back side of the input mirror (about 1 cm from the spot of the incident beam) and the second on the aluminum pilot model body itself. The minimal temperature reached at the model body was about 5 K while the mirrors itself were cooled down up to 14-16 K.

The two integral characteristics of the FP cavity were measured: finesse (sharpness) and the fraction of reflected light (contrast). The last parameter was defined as $(P_{ins} – P_{ref}) / P_{ins}$, where $P_{ins}$ – the incident laser power, and $P_{ref}$ – laser power reflected from the FP interferometer in the frequency-locked regime. Such value gives us an estimation of optical power taking part in the interference. The values mentioned have a direct influence to the sensitivity of opto - acoustical gravitational antenna.

It is known that a fraction of reflected light depends on the mode matching between the laser beam and the FP cavity. Thermal lens effect in mirrors destroys such matching and decrease the contrast. To check this effect for $CaF_2$ mirrors due to absorption light we have performed measurements of the part of reflected light at different laser powers, namely at 10, 20, 50, 100, 200 and 480 mW. For the low powers mirror's temperature measured by thermo sensor attached on the back side of the mirror and was practically the same as for the aluminum body. At the maximum power of 480 mW, the temperature difference between the mirrors and the aluminum body was about 14 - 15 K. The incident laser power was measured after reaching a quasi-stable point of thermal equilibrium, where the mirror's temperature was not changed during the measurement time. In the situations when the temperature changed relatively quickly, optical parameters was investigated only at minimal (10 mW) and maximum (480 mW) laser powers.

To control finesse evolution, we measured the width of FP cavity resonance peak by sweeping the laser frequency. Finesse mostly depends on reflectivity of the mirrors. We didn't find any significant changes of its value during cooling-heating cycles. It was close the same initial value of F=2300 for all available laser powers.

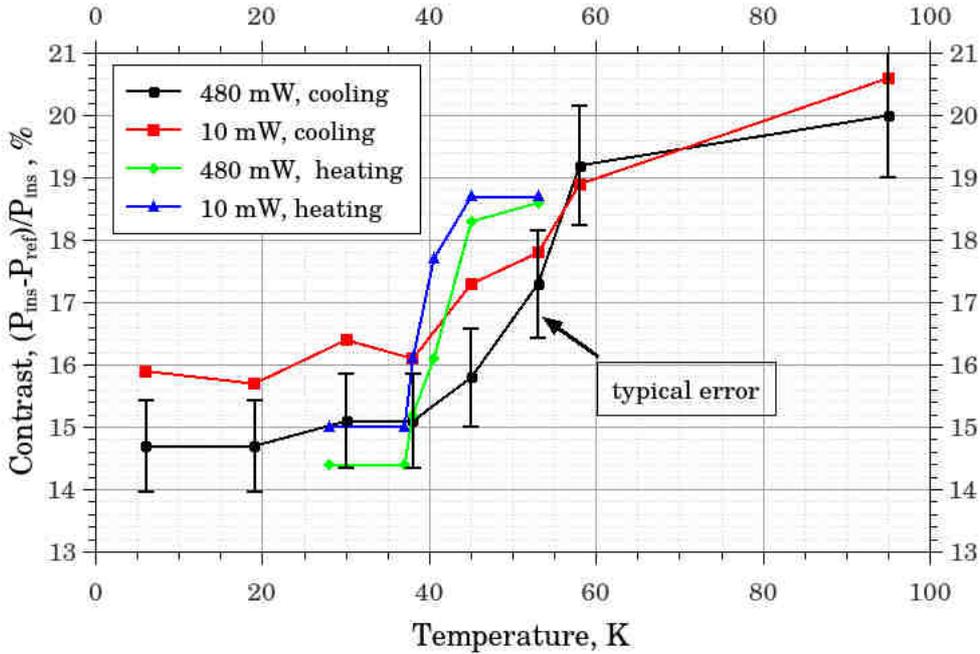

Fig.5. Contrast evolution for the FP-cavity versus temperature and different laser pump powers. Black and red curves present the cavity contrast during the cooling process with 480 mW and 10 mW powers. Green and blue curves give the contrast during the warming process.

The experimental data for the "contrast" measurement in the condition of FP-cavity optical resonance are presented in Fig. 5 (only data on minimal and maximal laser powers are presented). At temperatures between 60 and 40 K, contrast value quickly decreases and becomes more flat below 40 K. This zone corresponds to the $CaF_2$ thermal conductivity increasing during cooling process from 1 W/cm K at 60 K up to 10 W/cm K at 30 K [20]. So the thermally induced lens effect might be amplified in this zone and a violation of the mode matching between beam and FP-cavity mode is enhanced. Another curious result detected is that the contrast during cooling and heating phase has some hysteresis.

The contrast variation was relatively small and does not exceed 20%. It should not produce a significant influence on the sensitivity of the OGRAN cryo-version. From our experiments we consider that $CaF_2$ substrate mirrors can be used in such cryo-setups. Nevertheless it should be noted that necessary to make more detailed experiments with the mirrors with $CaF_2$ substrates as well as mirrors with sapphire substrates.

## 6. Discussion and conclusions.

Above we presented arguments in favor of a possibility of increasing the ORGAN gravitational detector sensitivity about two orders of magnitude due to its moderate cooling up to nitrogen temperature. Under this the expected strain spectral noise density has to reach $10^{-21}$ $Hz^{-1/2}$. It was shown that readout noises can be reduced to this level and even more if one will use the laser pump with 100 W instead of 1 W as it was taken in our estimates. The reason of considering such variant of modernization lays of course in a realistic view on a reconstruction of the existing OGRAN chamber [1,2] into nitrogen cryostat. For helium cryostat version it will require a construction of completely new setup very expensive in creation and exploiting [4,5].

In the process of this article preparation the first detection of GW-signal from black hole binary coalescence was declared [14] on LIGO detectors. It is interesting to note that the level of this signal amplitude was just close to $10^{-21}$ $Hz^{-1/2}$. Although the frequency range of the signal (~$10^2$ Hz) was much lower the OGRAN frequency. Unlike black holes, neutron stars during binary coalescence can generate gravitational waves in the kilohertz range. It gives us a hope that the level $10^{-21}$ $Hz^{-1/2}$ is enough perspective for the detection such type of events in our Galaxy and environment with radius ~20 Mpc.

In our final remarks we would like remind the famous case with neutrino and GW-bursts phenomenon from SN 1987A [15,16]. It was the unique example of search for "nu – gw" correlation between signals registered neutrino telescopes and room temperature gravitational wave detectors. Although only neutrino events have got recognition as "first registration" of neutrino from "collapsing star" [17], some algorithms of multichannel detection of "relativistic catastrophe" was proposed and tested. It was based at empirical statistics which presents stochastic coincidences rate between two channels of different nature under arbitrary relative time shifts of both date sets. The same method was applied in estimation of transient pulse statistics of LIGO detectors in [14].

The fact of "nu – gw" correlation was not confirmed [18,19] so as the sensitivity of room temperature bar antennas was insufficient for registration signals of astrophysical origin. The cryo-OGRAN will have sensitivity value at 4-5 orders of magnitude better then the room temperature bars detectors. The strategy of joint search for events of neutrino telescope (such as Baksan Neutrino Observatory of INR RAS - BUST) and gravitational wave detector (OGRAN) looks reasonable.

The crucial technology for realization of the cryo-OGRAN version is the technology of cryogenic mirrors. Although we have presented the positive experimental test with $CaF_2$ mirrors it might be considered only as a first preliminary step. The important point is a long term behavior of such mirrors during the long time observation series of the OGRAN as the gravitational wave antenna. Partly this problem could find solution in our planned experiments with the test setup.

## Acknowledgement

Authors would like to gratitude the academicians S.N. Bagaev and A.M. Cherepashchuk for attention to this work and stimulating discussions.
This work was supported in part by M.V. Lomonosov Moscow State University Program of Development as well as RFBR grants 14-02-00567 and 14-22-03036.